\documentclass[aps,prb,showpacs,twocolumn,floats]{revtex4}
\usepackage{graphicx}
\usepackage{subfigure}
\usepackage{epsfig}
\usepackage{dcolumn}
\usepackage{bm}

\begin{document}

\title{Rotational Doppler Effect in Magnetic Resonance}
\date{\today}
\author{S. Lend\'{i}nez$^{1,2}$, E. M. Chudnovsky$^{1,3}$, and J. Tejada$^{1,2}$}
\affiliation{$^{1}$Departament de F\'{i}sica Fonamental, Facultat
de F\'{i}sica, Universitat de Barcelona, Avinguda Diagonal 645,
08028 Barcelona, Spain\\ $^2$Institut de Nanoci\`{e}ncia i
Nanotecnologia IN2UB, Universitat de Barcelona, c. Mart\'{i} i
Franqu\`{e}s 1, 08028 Barcelona, Spain\\ $^{3}$Physics Department,
Lehman College, The City University of New York, 250 Bedford Park
Boulevard West, Bronx, NY 10468-1589, U.S.A.}
\date{\today}

\begin{abstract}
We compute the shift in the frequency of the spin resonance in a
solid that rotates in the field of a circularly polarized
electromagnetic wave. Electron spin resonance, nuclear magnetic
resonance, and ferromagnetic resonance are considered. We show
that contrary to the case of the rotating LC circuit, the shift in
the frequency of the spin resonance has strong dependence on the
symmetry of the receiver. The shift due to rotation occurs only
when rotational symmetry is broken by the anisotropy of the
gyromagnetic tensor, by the shape of the body, or by
magnetocrystalline anisotropy. General expressions for the
resonance frequency and power absorption are derived and
implications for experiment are discussed.
\end{abstract}

 \maketitle

\section{Introduction}
\begin{figure}
\vspace{-0.2cm} \hspace{2cm}
\includegraphics[width=90mm]{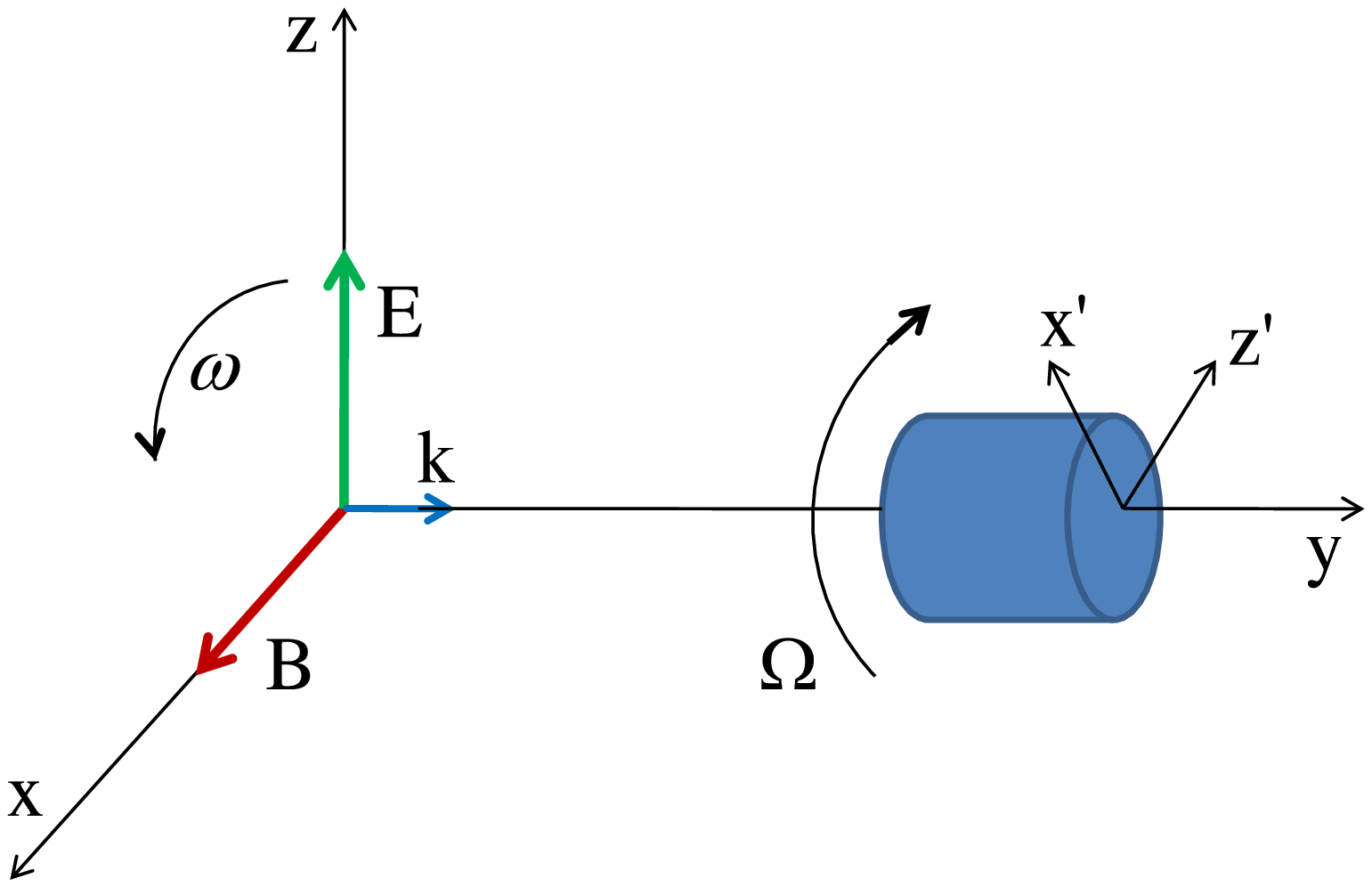}
\vspace{-2.5cm} \caption{Color online: Rotational Doppler effect.
The frequency $\omega$ of the circularly polarized electromagnetic
wave $(\omega,{\bf k})$ is the angular velocity of the rotation of
the electric (magnetic) field due to the wave at a given point in
space. The rotation of the receiver at an angular velocity
$\Omega$, depending on the direction of the rotation and the
helicity of the wave, adds or subtracts $\Omega$ to the frequency
of the wave $\omega$, rendering $\omega' = \omega \pm \Omega$ in
the coordinate frame of the receiver.} \label{RDE}
\end{figure}

The term Rotational Doppler Effect (RDE) is used to describe a
frequency shift encountered by a receiver of electromagnetic
radiation when either the receiver or the source of the radiation
are rotating. The effect is illustrated in Fig.\ \ref{RDE}. The
frequency of the wave, $\omega = 2\pi f$, measured at a given
point in space, corresponds to the angular velocity of the
rotation of the electric (magnetic) field due to the wave. If the
receiver is rotating mechanically at an angular velocity $\Omega$
about the axis parallel to the wave vector ${\bf k}$, than the
frequency of the wave perceived by the receiver equals
\begin{equation}\label{phase}
\omega' = \omega \pm \Omega\,.
\end{equation}
The sign, plus or minus, depends on the helicity of the wave and
the direction of the rotation of the receiver.

The RDE is less commonly known than the conventional Doppler
effect. One reason is that it is more difficult to observe.
M\"{o}ssbauer technique provides the most sensitive method for the
study of the frequency shift due to the conventional Doppler
effect, $\delta \omega = (v/c)\omega$ for $v \ll c$. The limiting
velocity has been a fraction of a millimeter per second and is due
to the finite very small linewidth of gamma radiation, $\delta
\omega/\omega \sim 10^{-13} - 10^{-12}$. Such a small linewidth
has even permitted observation of the transverse Doppler effect
\cite{Hay1960-Champeney1961-Kundig1963,Kholmetskii2008-Kholmetskii2009}
by performing M\"{o}ssbauer experiment on a rotating platform.
This effect, not to be confused with the RDE, consists of the
frequency shift $\delta \omega/\omega = -v^2/(2c^2)$ due to the
relativistic time dilation for a receiver moving tangentially with
respect to the source of the radiation. It is easy to see,
however, that the frequency shift as little as $\Omega/\omega \sim
10^{-13} - 10^{-12}$ due the RDE would require angular velocity of
the emitter or the receiver in the M\"{o}ssbauer experiment on the
order of a few MHz or even a few tens of MHz. The latter is still
one-two orders of magnitude greater than the angular velocities of
high-speed rotors used for magic angle spinning in NMR
applications.

The RDE frequency shift caused by a rotating plate inserted into a
beam of circularly polarized light was reported in Refs.
\onlinecite{Garetz1979,Simon1988,Bretaneker1990,Basstiy2002,Chen2008}.
The RDE was predicted for rotating light beams \cite{Nienhuis1996}
and subsequently observed using millimeter waves
\cite{Courtial1998} as well as in the optical range
\cite{Barreiro2006} (see Ref. \onlinecite{Padgett2006} for
review). In solid state experiments the RDE has proved
surprisingly elusive. Frequencies of the ferromagnetic resonance
(FMR) are typically in the GHz range or higher, which is far above
achievable angular velocities of mechanical rotation of
macroscopic magnets. However, small magnetic particles in beams
\cite{Xu-2005} or in nanopores \cite{FR} may rotate very fast.
Eq.\ (\ref{phase}) was recently applied to the analysis of the
observed anomalies in the FMR data on rotating nanoparticles
\cite{FR}. The RDE may be especially important for the NMR
technology that uses rapidly spinning samples. Frequency shifts of
the quadrupole line in the nuclear magnetic resonance (NMR)
experiment with a rotating sample were reported in Ref.
\onlinecite{Tycko1987} and analyzed in terms of Berry phase
\cite{Berry}. It was never fully explained, however, why such
shifts do not persist in the NMR experiments in which the angular
velocity of the magic-angle-spinning rotor with the sample often
exceeds the linewidth by an order of magnitude. Some hint to
answering this question can be found in Ref. \onlinecite{BB-1997}
that studied the effect of the rotation on radiation at the atomic
level. The authors of this work correctly argued that the RDE can
only be seen in the radiation of atoms and molecules placed in the
environment that destroys rotational symmetry.

Situation depicted in Fig.\ \ref{RDE} rather obviously leads to
the frequency shift by $\Omega$ when the emitter and the receiver
are based upon LC circuits. This has been tested by the GPS for
the case of a receiving antenna making as little as $8$
revolutions per second as compared to the carrier frequency of the
electromagnetic waves in the GHz range \cite{Ashby}. Eq.\
(\ref{phase}) has been also applied to the explanation of the
frequency shift encountered by NASA in the communications with
Pioneer spacecrafts \cite{Mashhoon-Pioneer}. One essential
difference between conventional and rotational Doppler effects is
that the first refers to the inertial systems while the second
occurs in the non-inertial systems. This prompted works that
considered RDE in the context of nonlocal quantum mechanics in the
accelerated frame of reference \cite{Mashhoon-Relativity}.
Relativity (or Galilean invariance for $v \ll c$) makes the
conventional Doppler effect quite universal. As we shall see
below, such a universality should not be expected for the RDE.
Indeed, the argument behind the RDE is based upon perception of a
circularly polarized wave by a rotating observer. Through the
Larmor theorem \cite{LL-FT} the mechanical rotation of the system
of charges is equivalent to the magnetic field. Consequently, when
making the argument, one has to check whether the resonant
frequency of the receiver is affected by the magnetic field.
Resonant frequencies of LC circuits are known to be insensitive to
the magnetic fields, thus making the argument rather solid. On the
contrary, the frequency of the receiver based upon magnetic
resonance would be sensitive to the fictitious magnetic field due
to rotation, thus making the argument incomplete.

In this paper we develop a rigorous theory of the RDE for magnetic
resonance. We show that the frequency shift due to rotation is
always different from $\Omega$. Broken rotational symmetry is
required for the shift to have a non-zero value, in which case the
magnetic resonance splits into two lines separated by $2\Omega$.
For the electron spin resonance (ESR) violation of the rotational
symmetry would naturally arise from the anisotropy of the
gyromagnetic tensor. In a solid state NMR experiment with a
rotating sample, violation of symmetry would be more common in the
presence of the magnetic order that provides anisotropy of the
hyperfine interaction. For a ferromagnetic resonance (FMR) the
asymmetry comes from the shape of the sample and from
magnetocrystalline anisotropy. The paper is organized as follows.
The physics of spin-rotation coupling is reviewed in Section
\ref{SR}. Frequency shift of the ESR in a rotating crystal with
anisotropic gyromagnetic tensor is computed in Section \ref{ESR}.
The effect of rotation on the NMR spectra is discussed in Section
\ref{NMR}. FMR in a rotating sample is studied in Section
\ref{FMRSec}. Power absorption by the rotating magnet is
considered in Section \ref{PowerSec}. Section \ref{Discussion}
contains some suggestions for experiment and discussion of
possible application of the RDE in solid state physics.

\section{Spin-rotation coupling}\label{SR}

In classical mechanics the Hamiltonian of the system in a rotating
coordinate frame is given by \cite{LL-M}
\begin{equation}\label{L}
{\cal{H}}' = \cal{H} - {\bf L}\cdot{\bf \Omega}\,.
\end{equation}
Here $\cal{H}$ is the Hamiltonian at $\Omega = 0$ and ${\bf L}$ is
the mechanical angular momentum of the system. For a system of
charges one can write
\begin{equation}
{\bf L} = \frac{\bf M}{\gamma}\,,
\end{equation}
where ${\bf M}$ is the magnetic moment and $\gamma$ is the
gyromagnetic ratio. Eq.\ (\ref{L}) then becomes equivalent to the
Hamiltonian,
\begin{equation}\label{M}
{\cal{H}}' = \cal{H} - {\bf M}\cdot{\bf B}\,,
\end{equation}
in the fictitious magnetic field,
\begin{equation}
{\bf B} = \frac{\bf \Omega}{\gamma}\,,
\end{equation}
which is the statement of the Larmor theorem \cite{LL-FT}.

Neither classical mechanics nor classical field theory deals with
the concept of a spin. The question then arises whether Eq.\
(\ref{L}) should contain spin ${\bf S}$ alongside with the orbital
angular momentum ${\bf L}$. Eq.\ (\ref{M}) hints that since the
magnetic moment can be of spin origin this should be the case.
Also it is known from relativistic physics that the generator of
rotations is
\begin{equation}
{\bf J} = {\bf L} + {\bf S}\,.
\end{equation}
It should be, therefore, naturally expected that in the presence
of a spin Eq.\ (\ref{L}) should be generalized as
\begin{equation}\label{LS}
{\cal{H}}' = \cal{H} - ({\bf L} + {\bf S})\cdot{\bf \Omega}\,.
\end{equation}
In quantum theory this relation can be rigorously derived in the
following way. Rotation by an angle ${\bm \phi}$ transforms the
Hamiltonian of an isolated system into \cite{QM}
\begin{equation}
\hat{\cal{H}}' =  \exp\left[\frac{i}{\hbar}
(\mathbf{L}+\mathbf{S})\cdot {\bm \phi
}\right]\hat{\cal{H}}\exp\left[-\frac{i}{\hbar}
(\mathbf{L}+\mathbf{S})\cdot {\bm \phi }\right]\,.
\end{equation}
To the first order on a small rotation ${\bm \phi}$ one obtains
\begin{equation}
\hat{\cal{H}}' =  \hat{\cal{H}} - \frac{i}{\hbar}
(\mathbf{L}+\mathbf{S})\cdot[\hat{\cal{H}},{\bm \phi}]\,,
\end{equation}
where we have taken into account that for an isolated system ${\bf
J}$ is conserved, that is ${\bf L} + {\bf S}$ commutes with
$\hat{\cal{H}}$. This equation becomes Eq.\ (\ref{LS}) if one
takes into account the quantum-mechanical relation
\begin{equation}
{\bf \Omega} = \frac{d {\bm \phi}}{dt} =
\frac{i}{\hbar}[\hat{\cal{H}},{\bm \phi}]
\end{equation}
and replaces operator ${\bf \Omega}$ by its classical expectation
value. For an electron Eq.\ (\ref{LS}) can be also formally
derived as a non-relativistic limit of the Dirac equation written
in the metric of the rotating coordinate frame \cite{Hehl1990}.
The answer for the corresponding Schr\"{o}dinger equation reads
\begin{equation}\label{Dirac}
i\hbar\frac{\partial \Psi}{\partial t} = \hat{\cal{H}}'\Psi\,,
\quad \hat{\cal{H}}' = \frac{{\hat{\bf p}}^2}{2m} -\left({\bf
r}\times \hat{\bf p} + \frac{1}{2}\hbar\hat{\bm
\sigma}\right)\cdot {\bf \Omega}\,,
\end{equation}
where ${\bf r}$ and ${\bf p} = -i\hbar{\bm \nabla}$ are the
radius-vector and the linear momentum of the electron,
respectively, and ${\sigma}_{x,y,z}$ are Pauli matrices.

There has been some confusion in literature regarding the term
$-{\bf S}\cdot {\bf \Omega}$ in the Hamiltonian of the body
studied in the coordinate frame that rotates together with the
body \cite{EC-PRL1994,Villain,CGS}. To elucidate the physical
meaning of this term, let us consider the resulting equation of
motion for a classical spin-vector \cite{CT}
\begin{equation}\label{LL-eq}
\frac{d{\bf S}}{dt} = -{\bf S} \times \frac{\delta
{\cal{H}}'}{\delta {\bf S}}\,.
\end{equation}
If $\cal{H}$ does not depend on spin, then the spin cannot be
affected in any way by the rotation of the body. In this case
${\delta {\cal{H}}'}/{\delta {\bf S}}=-{\bf \Omega}$ and Eq.\
(\ref{LL-eq}) simply describes the precession of ${\bf S}$ about
${\bf \Omega}$:
\begin{equation}
\frac{d{\bf S}}{dt} = {\bf S} \times {\bf \Omega}\,.
\end{equation}
It shows how a constant vector ${\bf S}$ (or any other vector to
this matter) is viewed by an observer rotating at an angular
velocity ${\bf \Omega}$. This has nothing to do with the
spin-orbit or any other interaction. Such interactions should be
accounted for in the $\hat{\cal{H}}$ part of the Hamiltonian
$\hat{\cal{H}}'$. The effect of rotations on various magnetic
resonances is considered in the next sections.
\\

\section{Frequency Shift of the Electron Spin Resonance Due to
Rotation}\label{ESR}

In this Section we consider an electron in a rotating crystal or
in a rotating quantum dot characterized by the anisotropic
gyromagnetic tensor, $g_{ij}$. The effect of local rotations due
to transverse phonons on the width of the ESR has been studied in
Ref.\ \onlinecite{CCG}. Here we are interested in the effect of
the global rotation on the ESR frequency. To deal with the
stationary states we shall assume that the axis of rotation ${\bf
\Omega}$ is parallel to the applied magnetic field ${\bf B}$ and
will compute the energy levels of the electron as measured by the
observer rotating together with the system. In the rotating frame
the spin Hamiltonian of the electron is
\begin{equation}\label{H-ESR}
\hat{\cal{H}}'= \frac{1}{2}\mu _{B}\,g_{ij}\,{\sigma}_{i}B_{j} -
\frac{1}{2}\hbar{\bm \sigma}\cdot{\bf \Omega}\,. \vspace{1mm}
\end{equation}
Positive sign of the first (Zeeman) term is due to the negative
gyromagnetic ratio $\gamma$ for the electron ($\mu_B = \hbar
|\gamma|$ being the Bohr magneton).

\begin{figure}
\vspace{0cm} \hspace{2cm}
\includegraphics[width=70mm]{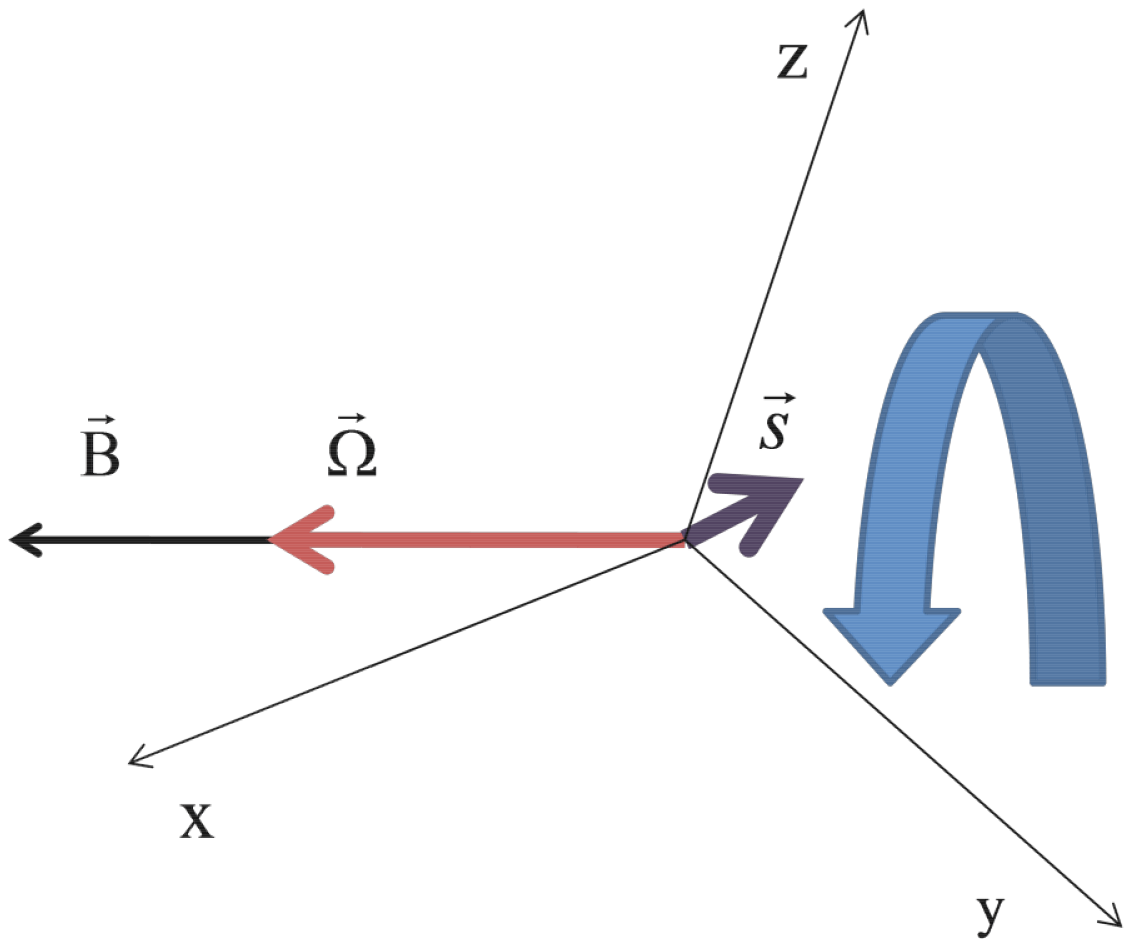}
\vspace{0cm} \caption{Color online: Spin in the magnetic field
parallel to the rotation axis of the crystal. The rotating
coordinate axes $x,y,z$ are chosen along the principal axes of the
gyromagnetic tensor.} \label{spin}
\end{figure}

The geometry of the problem is illustrated in Fig.\ \ref{spin}. In
the rotating frame the solid matrix containing the electron is
stationary. It is convenient to choose the coordinate axes of that
matrix along the principal axes of the tensor $g_{ij}$. Then
$g_{ij}$ is diagonal,
\begin{equation}
g_{ij}=g_{i}\delta _{ij}, \label{gdiag}
\end{equation}
represented by three numbers, $g_{x}$, $g_{y}$, and $g_{z}$ that
can be directly measured when the system is at rest. Eq.\
(\ref{H-ESR}) then becomes
\begin{eqnarray}\label{diagonal}
\hat{\cal{H}}' &= &\frac{1}{2}\left[(\mu _{B}\,g_{x}B_x
-\hbar\Omega_x)\,{\sigma}_{x} + (\mu _{B}\,g_{y}B_y
-\hbar\Omega_y)\,{\sigma}_{y} \right. \nonumber \\
& + & \left.(\mu _{B}\,g_{z}B_z
-\hbar\Omega_z)\,{\sigma}_{z}\right]\,.
\end{eqnarray}
Diagonalization of this Hamiltonian with the account of the fact
that ${\bf \Omega}$ was chosen parallel to ${\bf B}$ gives the
following energy levels of $\hat{\cal{H}}'$:
\begin{equation}\label{levels}
E_{\pm} = \pm\frac{1}{2}\mu_B B\left[ \sum_{i=x,y,z}\left(g_{i
}-\frac{\hbar \Omega}{\mu_B B}\right)^{2}n_{i}^{2}\right] ^{1/2}\,
\end{equation}
Here ${\bf n}$ is the unit vector in the direction of the axis of
rotation,
\begin{equation}
{\bf n} = \frac{\bf \Omega}{\Omega} = \frac{\bf B}{B}\,.
\end{equation}

In practice, the angular velocity of the mechanical rotation will
always be sufficiently small to provide the condition $\hbar\Omega
\ll \mu_B B$. Contribution of the rotation to the ESR frequency in
the rotating frame,
\begin{equation}
\hbar\omega'_{ESR} = {E_+ - E_-}\,,
\end{equation}
will, therefore, be small compared to the ESR frequency
\begin{equation}
\hbar\omega_{ESR} =
{\mu_B}B(g_{x}^{2}n_{x}^{2}+g_{y}^{2}n_{y}^{2}+g_{z}^{2}n_{z}^{2})
^{1/2}
\end{equation}
unperturbed by rotation. Expanding Eq.\ (\ref{levels}) to the
first order in $\Omega$ one obtains
\begin{eqnarray}
&& \omega'_{ESR} =  \omega_{ESR} - \kappa \Omega\,, \\
\label{ESR-shift} && \kappa  = \frac{g_xn_x^2 + g_yn_y^2 +
g_zn_z^2}{\sqrt{g_x^2n_x^2 + g_y^2n_y^2 + g_z^2n_z^2}}
\label{ESR-kappa}\,.
\end{eqnarray}
Here $\Omega$ can be positive or negative depending on the
direction of rotation.

Few observations are in order. Firstly, according to Eq.\
(\ref{ESR-kappa}), the frequency shift for the observer rotating
together with the sample containing the electron is never zero.
Secondly, when the rotation is about one of the principal axes of
the gyromagnetic tensor, Eq.\ (\ref{ESR-kappa}) gives $\kappa =
1$, so that the frequency shift for the rotating observer is
exactly $\Omega$. The ESR occurs when the frequency $\omega'$ of
the circularly polarized electromagnetic wave perceived by the
rotating observer and given by Eq.\ (\ref{phase}) coincides with
$\omega'_{ESR}$. If the rotation is about one of the principal
axes of $g_{ij}$, then $\kappa =1$ and the angular velocity
$\Omega$ cancels exactly from the equation $\omega' =
\omega_{ESR}'$ for the polarization of the wave that corresponds
to $\omega' = \omega - \Omega$, thus, resulting in no RDE
frequency shift for an experimentalist working in the laboratory
frame. For the opposite polarization of the wave, corresponding to
$\omega' = \omega + \Omega$, the shift in the rotationally
invariant case formally equals $2\Omega$. However, such photons
would have their spin projection in the direction opposite to the
one necessary to produce the spin transition. They can be absorbed
only when the rotational symmetry is broken so that the electron
spin in the direction of the wave vector is no longer a good
quantum number (see Section \ref{PowerSec}).

\section{Frequency Shift of the Nuclear Magnetic Resonance Due to
Rotation}\label{NMR}

Let us consider a nuclear spin ${\bf I}$ in the magnetic field
parallel to the axis of rotation of the sample. It is clear from
the previous section that the mechanical rotation combined with
the rotationally invariant Zeeman interaction of the nuclear
magnetic moment with the field,
\begin{equation}\label{H-NMR}
\hat{\cal{H}}'= -\gamma_n g_n{\bf I}\cdot{\bf B} - {\bf
I}\cdot{\bf \Omega}\,,
\end{equation}
(with $\gamma_n > 0$ and $g_n$ being nuclear gyromagnetic ratio
and gyromagnetic factor, respectively) are not sufficient to
produce the RDE. Isotropic hyperfine interaction with an atomic
spin ${\bf S}$ of the form $-A{\bf I}\cdot{\bf S}$ would not
change this either. However, an anisotropic hyperfine interaction,
\begin{equation}\label{HF}
\hat{\cal{H}}_{hf} = -A_{ij}I_iS_j\,,
\end{equation}
in principle, can do the job. If there is a ferromagnetic order in
the solid, then ${\bf S}$ develops a non-zero average, $\langle
{\bf S} \rangle$. Replacing $S_j$ in Eq.\ (\ref{HF}) with $\langle
S_j \rangle$ and adding the hyperfine interaction to Eq.\
(\ref{H-NMR}), one obtains
\begin{equation}\label{NMR-HF}
\hat{\cal{H}}'= -\gamma_n g_n{\bf I}\cdot{\bf B} -
A_{ij}I_i\langle S_j\rangle - {\bf I}\cdot{\bf \Omega}\,.
\end{equation}

To work with the stationary energy states in the rotating frame,
we shall assume that all three vectors ${\bf B}$, $\langle {\bf S}
\rangle$, and ${\bf \Omega}$ are parallel to each other. Let us
study the case of $I = 1/2$. Choosing the coordinate axes along
the principal axes of tensor $A_{ij}=A_i\delta_{ij}$, it is easy
to see that Eq.\ (\ref{NMR-HF}) is equivalent to the Zeeman
Hamiltonian,
\begin{equation}\label{NMR-eff}
\hat{\cal{H}}' =
-\frac{1}{2}\mu_n\left[g_x^{eff}\sigma_xB_x+g_y^{eff}\sigma_yB_y+g_z^{eff}\sigma_zB_z\right]
\end{equation}
with an effective gyromagnetic tensor whose principal values are
given by ($i=x,y,z$)
\begin{equation}
g_i^{eff} = g_n + \frac{B^{hf}_i}{B} + \frac{\hbar\Omega}{\mu_n
B}\,, \vspace{1mm}
\end{equation}
where we have introduced the nuclear magneton, $\mu_n = \hbar
\gamma_n$, and the hyperfine field, ${\bf B}^{hf}$, with
components
\begin{equation}
B^{hf}_i =\frac{\hbar A_i |\langle{\bf S}\rangle|}{\mu_n}\,.
\end{equation}
The energy levels of the Hamiltonian (\ref{NMR-eff}) are
\begin{equation}\label{NMR-levels}
E_{\pm} = \pm\frac{1}{2}\mu_n B\left[ \sum_{i=x,y,z}\left(g_{i
}^{eff}\right)^{2}n_{i}^{2}\right] ^{1/2}\,,
\end{equation}
where ${\bf n} = {\bf B}/{B}$.

Let us consider the case of small $\Omega$. Making the series
expansion of Eq.\ (\ref{NMR-levels}) one obtains to the first
order on $\Omega$
\begin{equation}\label{NMR-freq}
\omega'_{NMR} = \frac{E_+-E_-}{\hbar} = \omega_{NMR} + \kappa
\Omega
\end{equation}
with $\kappa$  given by
\begin{equation}\label{NMR-kappa}
\kappa = \frac
{\sum_{i=x,y,z}\left(g_n+{B^{hf}_i}/{B}\right)n_i^2}
{\sqrt{\sum_{i=x,y,z}\left(g_n+{B^{hf}_i}/{B}\right)^2n_i^2}}\,.
\end{equation}
In the case of the isotropic hyperfine interaction, $B^{hf}_x =
B^{hf}_y=B^{hf}_z$ (that is, $A_x=A_y=A_z$), Eq.\
(\ref{NMR-kappa}) gives $\kappa = 1$. Same situation occurs when
the direction of the field and the axis of rotation coincide with
one of the principal axes of the tensor of hyperfine interactions.
For arbitrary rotations Eq.\ (\ref{NMR-kappa}) gives $\kappa
\rightarrow 1$ when $B \gg B^{hf}$, making the frequency shift
defined by $\omega' = \omega'_{NMR}$ negligible for the
polarization ($\omega' = \omega + \Omega$) that is predominantly
absorbed due to the selection rule. Is is likely, therefore, that
a significant RDE in the NMR can be observed only in magnetically
ordered materials, in the field comparable or less than the
hyperfine field, for rotations about axes that do not coincide
with the symmetry axes of the crystal. If these conditions are
satisfied, and the width of the resonance is not very large
compared to $\Omega$, the NMR produced by linearly polarized waves
would split into two lines of uneven intensity separated by
$2\Omega$. In fact, the existing experimental techniques permit
observation of this effect (see Section \ref{Discussion}).

\section{Frequency Shift of the Ferromagnetic Resonance Due to
Rotation}\label{FMRSec}

We now turn to the rotating ferromagnets. We begin with a simplest
model of ferromagnetic resonance studied by Kittel \cite{Kittel}.
In this model one neglects the effects of magnetocrystalline
anisotropy and considers a uniformly magnetized ferromagnetic
ellipsoid in the external magnetic field ${\bf B} = \mu_0 {\bf H}$
(with $\mu_0$ being the magnetic permeability of vacuum). The
energy density of such a ferromagnet is determined by its Zeeman
interaction with the external field and by magnetic dipole-dipole
interactions inside the ferromagnet:
\begin{equation}\label{H-FM}
{\cal{H}} = \mu_0\left[- {\bf M}\cdot{\bf H} +
\frac{1}{2}N_{ij}M_iM_j\right]\,.
\end{equation}
Here ${\bf M}$ is the magnetization and $N_{ij}$ is tensor of
demagnetizing coefficients. The principal axes of $N_{ij}$
coincide with the axes of the ellipsoid. Choosing the coordinate
axes along the principal axes and taking into account that for a
ferromagnet
\begin{equation}
{\bf M}^2 = M_x^2 + M_y^2 + M_z^2 = M_0^2
\end{equation}
is a constant, one can rewrite Eq.\ (\ref{H-FM}) as
\begin{equation}\label{H-FM-xyz}
{\cal{H}} = - \mu_0\left[{\bf M}\cdot{\bf H} + \frac{1}{2}(N_x -
N_z)M_x^2 + \frac{1}{2}(N_y - N_z)M_y^2)\right],
\end{equation}
where we have omitted unessential constant. For, e.g., an infinite
circular cylinder $N_x=N_y=1/2, N_z=0$. In general, for an
ellipsoid elongated along the $Z$-axis one has $N_x-N_z > 0,
N_y-N_z >0$, so that in the absence of the field the minimum of
Eq.\ (\ref{H-FM-xyz}) corresponds to ${\bf M}$ in the
$Z$-direction. This will still be true in the external field if
the latter is applied in the $Z$-direction, which is the case we
consider here. Note that a finite field is always needed to
prevent the magnet from breaking into magnetic domains.

The FMR frequency, $\omega_{FMR}$, can be obtained from either
classical or quantum mechanical treatment \cite{CT}. Classically,
it is the frequency of the precession of ${\bf M}$ about its
equilibrium direction. To find $\omega_{FMR}$ one should linearize
the equation,
\begin{equation}
\frac{d{\bf M}}{dt} = \gamma{\bf M} \times {\bf B}^{(eff)}\,,
\quad {\bf B}^{(eff)} = - \frac{\delta \cal{H}}{\delta {\bf M}}\,,
\end{equation}
around ${\bf M} = M_0{\bf e}_z$ ($\gamma < 0$ being the
gyromagnetic ratio). The answer reads \cite{Kittel}
\begin{equation}\label{FMR}
\omega_{FMR} = \sqrt{\omega_x \omega_y}\,,
\end{equation}
where
\begin{eqnarray}\label{omega-xy}
\omega_{x} & = & |\gamma|[B + (N_{x}-N_z)\mu_0 M_0] \nonumber \\
\omega_{y} & = & |\gamma|[B + (N_{y}-N_z)\mu_0M_0]\,.
\end{eqnarray}

To study the RDE we should now solve the same problem in the
coordinate frame rotating about the $Z$-axis at an angular
velocity $\Omega$. In the presence of rotation the Hamiltonian
becomes
\begin{equation}\label{H-rot}
{\cal{H}}' = \cal{H} -\frac{\bf M}{\gamma}\cdot{\bf \Omega}\,.
\end{equation}
It is easy to see that for ${\bf \Omega} = \Omega{\bf e}_z$ this
effectively adds ${\bf \Omega}/\gamma$ to the external field.
Consequently, the FMR frequency in the rotating frame becomes
\begin{equation}\label{FMR-rot}
\omega_{FMR}'  = \sqrt{\omega'_x \omega'_y}
\end{equation}
with
\begin{eqnarray}\label{omega-rot-xy}
\omega'_{x} & = & |\gamma|\left[B + \frac{\Omega}{\gamma} + (N_{x}-N_z)\mu_0M_0\right] \nonumber \\
\omega'_{y} & = & |\gamma|\left[B +  \frac{\Omega}{\gamma} +
(N_{y}-N_z)\mu_0 M_0\right]\,.
\end{eqnarray}

Our immediate observation is that for a symmetric ellipsoid ($N_x
= N_y$)
\begin{equation}
\omega_{FMR}' = \omega_{FMR} - \Omega\,,
\end{equation}
so that the RDE frequency shift determined by the equation
$\omega' = \omega - \Omega = \omega_{FMR}'$ is exactly zero. For
an asymmetric ellipsoid ($N_x \neq N_y$), expanding Eq.\
(\ref{FMR-rot}) into a series on $\Omega$ one obtains to the first
order
\begin{equation}
\omega_{FMR}' = \omega_{FMR} - \kappa\Omega\,,
\end{equation}
with
\begin{equation}\label{kappa-FMR}
\kappa = \frac{1}{2}\left(\sqrt{\frac{\omega_x}{\omega_y}} +
\sqrt{\frac{\omega_y}{\omega_x}}\right)\,.
\end{equation}
It is easy to see that $\kappa \geq 1$. At large fields, $B \gg
\mu_0M_0$, equations (\ref{omega-xy}) and (\ref{kappa-FMR}) give
$\kappa \rightarrow 1$, that is, no frequency shift due to the
RDE. Sizable frequency shift of the FMR observed in the laboratory
frame due to the rotation of the sample should occur only at $B$
not significantly exceeding $\mu_0M_0$ and only in a sample
lacking the rotational symmetry.

One can easily generalize the above approach to take into account
any type of the magnetocrystalline anisotropy. The formulas look
especially simple in the case of the second-order anisotropy. Such
anisotropy adds the term
\begin{equation}
-\frac{1}{2}\mu_0\beta_{ij}M_iM_j
\end{equation}
to the Hamiltonian of the magnet, with $\beta_{ij}$ being some
dimensionless symmetric tensor. Consider, e.g., an orthorhombic
crystal whose axes ($a,b,c$) coincide with the axes of the
ellipsoid and whose easy magnetization axis, $c$, is parallel to
the $Z$-direction. In this case all the above formulas remain
valid if one replaces the demagnetizing factors with
\begin{equation}
N_i' = N_i - {\beta_i}\,, \quad i = x,y,z \,,
\end{equation}
where $\beta_{x}$, $\beta_{y}$, and $\beta_{z}$ are the principal
values of $\beta_{ij}$. Due to the orthorhombic anisotropy ($a
\neq b \rightarrow \beta_x \neq \beta_y$) the RDE may now occur
even in a sample of the rotationally invariant shape ($N_x=N_y$).

\section{Power Absorption by a Rotating Magnet}\label{PowerSec}

\begin{figure} \vspace{0cm}
\hspace{2cm}
\includegraphics[width=80mm]{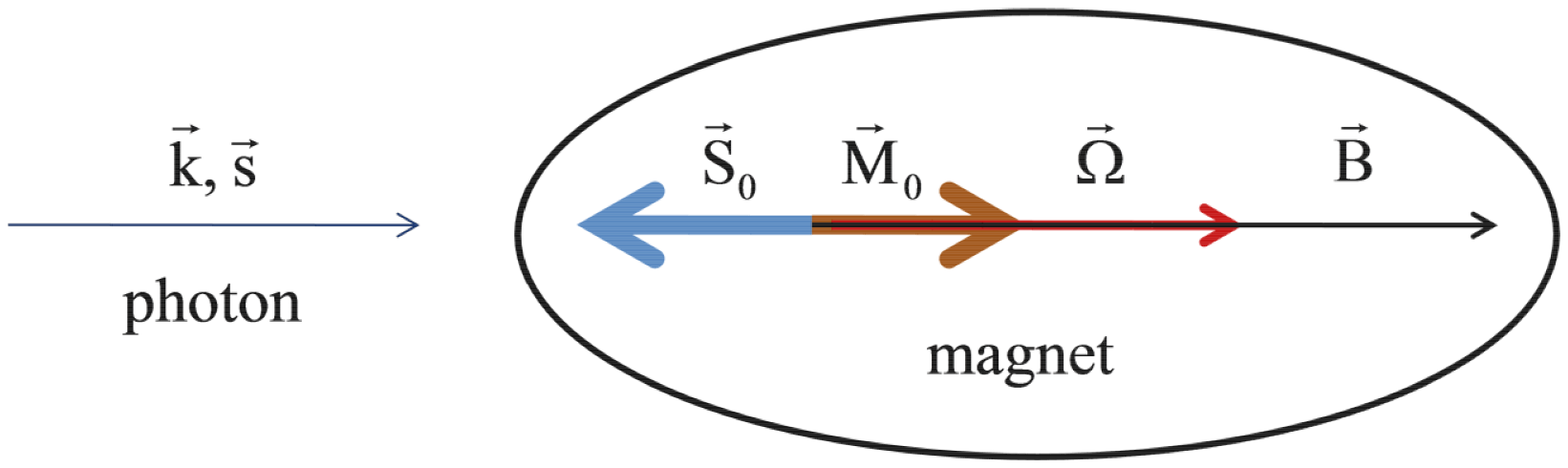}
\vspace{0cm} \caption{Color online: Geometry of the FMR studied in
the paper. Ferromagnet uniformly magnetized by a static magnetic
field, ${\bf B}$, is rotating at an angular velocity ${\bf
\Omega}$ in the radiation field of circularly polarized photons of
wave vector ${\bf k}$ and spin ${\bf s}$. (Due to the negative
gyromagnetic ratio, the equilibrium spin of the magnet, ${\bf
S}_0$, is antiparallel to its equilibrium magnetic moment ${\bf
M}_0$.)} \label{magnet}
\end{figure}

For non-relativistic rotations the radiation power absorbed by the
magnet should be the same in the laboratory frame and in the
rotating frame. Calculation in the rotating frame is easier. We
shall assume that the dimensions of the sample are small compared
to the wavelength of the radiation, so that the field of the wave
at the position of the ferromagnet is nearly uniform. The geometry
studied below is illustrated in Fig.\ \ref{magnet}.
Within the model of Eq.\ (\ref{H-rot}), the rotating magnet placed
in the field of a circularly polarized wave feels the oscillating
magnetic field that can be represented by a complex function
\begin{equation}\label{cp-wave}
h(t) = h_0 e^{\pm i\omega't}\,, \quad \omega' = \omega \mp \Omega
\end{equation}
giving the components of the field as
\begin{equation}
h_x = {\rm Re} (h), \; h_y = {\rm Im} (h)\,.
\end{equation}
Here $h_0$ is the complex amplitude of the wave, $\pm$ sign in
Eq.\ (\ref{cp-wave}) determines the helicity of the wave, while
the sign of $\Omega$ determines the direction of rotation of the
magnet. Due to the wave the magnetization acquires a small
ac-component $m(t)$ (whose real and imaginary parts represent
$m_x$ and $m_y$, respectively),
\begin{equation}
m(t) = \hat{\chi}(\omega)h(t)\,,
\end{equation}
where $\hat{\chi}$ is the susceptibility tensor. The absorbed
power is given by \cite{CT}
\begin{equation}
P = \pm i\mu_0\omega'h_0^*(\hat{\chi} -
{\hat{\chi}}^{\dagger})h_0\,.
\end{equation}
The problem has, therefore, reduced to the computation of the
susceptibility in the rotating frame. The latter can be done by
solving the Landau-Lifshitz equation,
\begin{equation}\label{LL-M}
\frac{d{\bf M}}{dt} = \gamma{\bf M} \times {\bf
B}^{(eff)}-\frac{\eta}{M_0}|\gamma|{\bf M}\times\left[{\bf M}
\times {\bf B}^{(eff)}\right]\,,
\end{equation}
in the rotating frame, that is, with ${\bf B}^{(eff)} = - {\delta
{\cal{H}}'}/{\delta {\bf M}}$ and
\begin{equation}
{\cal{H}}' = \cal{H} -\frac{\bf M}{\gamma}\cdot{\bf \Omega} - {\bf
M}\cdot{\bf h}\,.
\end{equation}
The parameter $\eta$ in Eq.\ (\ref{LL-M}) is a dimensionless
damping coefficient that is responsible for the width of the FMR
in the absence of inhomogeneous broadening.

Substituting ${\bf M} = M_0{\bf e}_z + {\bf m}$ into Eq.\
(\ref{LL-M}) and solving for $\hat{\chi}$ one obtains for the
power
\begin{equation}\label{power}
P_{\pm} = \frac{1}{2}\eta |\gamma| M_0 \mu_0^2
|h_0|^2f_{\pm}(\omega')\,,
\end{equation}
where
\begin{equation}\label{cp-f}
f_{\pm} = \frac{{\omega'}^2[2({\omega'}^2 - {\omega'}_{FMR}^2) \pm
2{\omega'}(\omega'_{x}+\omega'_{y})
+(\omega'_{x}+\omega'_{y})^2]}{({\omega'}^2 - {\omega'}_{FMR}^2)^2
+ \eta^2 \omega'^2(\omega'_{x} + \omega'_{y})^2}\,.
\end{equation}
Notice that when there is a full rotational symmetry,
$\omega'_{x}=\omega'_{y}=\omega'_{FMR}$, the absorbed power at the
resonance is non-zero only for one polarization of the wave that
corresponds to the upper sign in Eqs.\ (\ref{cp-wave}) and
(\ref{cp-f}). This is a consequence of the selection rule due to
conservation of the $Z$-component of the total angular momentum
(absorbed photon + excited magnet).

Let us now consider a rotating ferromagnet in the radiation field
of a linearly polarized electromagnetic wave. In the rotating
frame the complex magnetic field of such a wave is
\begin{equation}
h(t) = \frac{h_0}{2}\left[e^{i(\omega - \Omega)t} +
e^{-i(\omega+\Omega)t}\right]=h_0e^{-i\Omega t} \cos (\omega t)
\end{equation}
Repeating the above calculation, one obtains for the power
averaged over the period of rotation
\begin{equation}
P = \frac{1}{8} \eta |\gamma| M_0 \mu_0^2 |h_0|^2
\left[f_{+}(\omega-\Omega) + f_{-}(\omega+\Omega)\right]\,.
\end{equation}
When the rotational symmetry of the magnet is broken, $\omega_x
\neq \omega_y$, $\kappa > 1$, the absorption has two maxima of
uneven height at
\begin{equation}
\omega = \omega_{FMR} - (\kappa \mp 1)\Omega\,.
\end{equation}
As the rotational symmetry is gradually restored, $\omega_x
\rightarrow \omega_y$, $\kappa \rightarrow 1$, the rotational
shift in the position of the main maximum disappears. In that
limit the shift in the position of a smaller maximum approaches
$2\Omega$ while the height of that maximum goes to zero, see Fig.\
\ref{power}.
\begin{figure} \vspace{0cm}
\hspace{2cm}
\includegraphics[width=80mm]{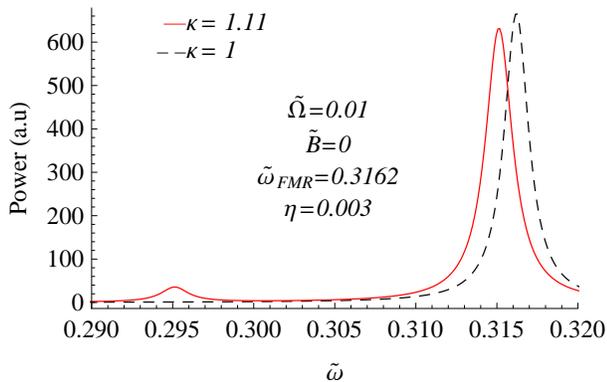}
\vspace{0cm} \caption{Color online: Absorption of power of
linearly polarized electromagnetic radiation by a rotating magnet.
Frequencies are given in the units of $\gamma \mu_0 M_0$. As the
rotational symmetry is violated the FMR becomes shifted and the
second FMR line emerges separated by $2\Omega$ from the first
line.} \label{power}
\end{figure}

\section{Discussion}\label{Discussion}

We have computed the frequency shift of the magnetic resonance due
to rotation of the sample. The effect of rotation on the ESR, NMR,
and FMR has been studied. We found that it is, generally, quite
different from the rotational Doppler effect reported in other
systems \cite{Padgett2006}. The differences stem from the
observation that the spin of an electron or an atom would be
insensitive to the rotation of the body as whole if not for the
relativistic spin-orbit coupling. Even with account of spin-orbit
interactions the spin would not simply follow the rotation of the
body but would exhibit more complex behavior described by the
dynamics of the angular momentum. Everyone who watched the
behavior of a gyroscope in a rotating frame would easily
appreciate this fact.

We found the following common features of the magnetic resonance
in a rotating sample.
\begin{itemize}
\item
If the spin Hamiltonian is invariant with respect to the rotation,
then the rotation of the body has no effect on the frequency of
the resonant absorption of a circularly polarized electromagnetic
wave.
\item
As the rotational invariance is violated, the absorption line
shifts. The shift is different from the angular velocity of
rotation, $\Omega$. It depends on the degree of violation of the
rotational symmetry. The frequency shift goes to zero when the
symmetry is restored.
\item
In the case of a linearly polarized radiation a second resonance
line emerges, separated by $2\Omega$ from the first line. The
intensity of that line depends on the degree of violation of
rotational symmetry. It disappears when the rotational symmetry is
restored.
\end{itemize}

ESR and FMR measurements are usually performed in the GHz range,
with the width of the resonance being sometimes as low as a few
MHz. Currently available small mechanical rotors can rotate as
fast as $100$kHz, which, nevertheless, is still low compared to
the linewidths of ESR and FMR. Note, however, that the position of
the ESR or FMR maximum can be determined with an accuracy of a few
hundred kHz. It is then not out of question that under appropriate
conditions the RDE frequency shift and the splitting of the
resonance can be observed in high precision ESR and FMR
experiments even when the rotation frequency is significantly
lower than the linewidth. Since anisotropy of the sample is needed
to provide rotational asymmetry, the measurements should be
performed on single crystals. Crystals with significant anisotropy
of the gyromagnetic tensor should be selected for ESR experiments.
When the magnetocrystalline anisotropy is weak, the RDE in FMR can
be induced by the asymmetric shape of the sample alone due to the
anisotropy of dipole-dipole interactions. Even in this case,
however, a single crystal would be preferred to provide a narrow
linewidth. Same applies to experiments on RDE in solid state NMR.
The NMR frequency range is much lower than that used in ESR and
FMR experiments. The width of the NMR line can be as low as a few
kHz, that is, well below the available rotational angular
velocities. The key to the observation of RDE in a solid state NMR
must be the use of a crystal having magnetic order and strong
anisotropy of the hyperfine interaction.

A separate interesting question is magnetic resonance in small
magnetic particles that are free to rotate. Particles of size in
the nanometer range can easily be excited into rotational states
with $\Omega$ of hundreds of MHz. Contrary to the rotational
quantum states of molecules that have been studied for decades,
analytical solution of the problem of a quantum-mechanical rotator
does not exist even without a spin. Presence of the spin
interacting with a mechanical rotation complicates this problem
even further. Rigorous solution has been recently found for the
low energy states of a rotator that can be treated as a two-state
spin system \cite{CG-2010}. General solution is very difficult to
obtain. In the case when a particle consists of a large number of
atoms, one can develop a semiclassical approximation in which
${\bf \Omega}$ is replaced with ${\bf L}/I$ (with $I$ being the
moment of inertia). This suggests that the magnetic resonance in
nanoparticles that are free to rotate would split into many lines
related to the quantization of ${\bf L}$. Some evidence of this
effect has been recently found in the FMR studies of magnetic
particles in nanopores \cite{FR}. Rapid progress in measurements
of single magnetic nanoparticles \cite{Wern-08} may shed further
light on their quantized rotational states and related spin
resonances.

\section{Acknowledgements}
S.L. acknowledges financial support from Grupo de
Investigaci\'{o}n de Magnetismo de la Universitat de Barcelona.
The work of E.M.C. has been supported by the grant No. DMR-0703639
from the U.S. National Science Foundation and by Catalan ICREA
Academia. J.T. acknowledges financial support from ICREA Academia.

\end{document}